# Blue to Near-IR Integrated PZT Silicon Nitride Modulators for Quantum and Atomic Applications

NICK MONTIFIORE,[1] ANDREI ISICHENKO,[1] NITESH CHAUHAN, [1,2,3] JIAWEI WANG,[1,4] ANDREW S. HUNTER,[1] MARK W. HARRINGTON,[1] RAHUL CHAWLANI[1], RYAN Q. RUDY,[5] IAIN KIERZEWSKI,[5] MICHAEL PUSHKARSKY,[6] AND DANIEL J. BLUMENTHAL[1,*]

[1]*Department of Electrical and Computer Engineering, University of California Santa Barbara, Santa Barbara, California 93106, USA*
[2]*Present Address: Time and Frequency Division, National Institute of Standards and Technology, Boulder, CO, 80305, USA*
[3]*Present Address: Department of Physics, University of Colorado, Boulder, Boulder, CO, 80309, USA*
[4]*Present Address: Lightmatter, Inc., Boston, MA, 02210, USA*
[5]*U.S. Army Combat Capabilities Development Command Army Research Laboratory, Adelphi, MD, 20783, USA*
[6]*DRS Daylight Solutions, San Diego, CA, 92127, USA*
*danb@ucsb.edu*

**Abstract:** Modulation and control of lasers and optical signals is necessary for trapped-ion and cold neutral atom quantum systems. Given the diversity of atomic species, experimental modalities, and architectures, integrated optical modulators that are designed to operate across the visible to near-infrared (NIR) spectrum are a key step towards portable, robust, and compact quantum computers, clocks, and sensors. Integrated optical modulators that are wavelength-independent, CMOS-compatible, and capable of maintaining low waveguide losses and a high resonator quality factor (Q), DC-coupled broadband frequency response, and low power consumption, are essential for scalable photonic integration. Yet progress towards these goals has remained limited. To demonstrate the versatility of this platform we demonstrate four types of integrated stress-optic lead zirconate titanate (PZT) silicon nitride ($Si_3N_4$) modulators - a coil Mach-Zehnder modulator, a coil pure phase modulator, and bus-coupled and add-drop ring resonator modulators, with operation from 493 nm to 780 nm. The PZT-actuated coil MZM operates at 532 nm with a $V\pi$ of 2.8V, a DC to 0.4 MHz 3-dB bandwidth, and an extinction ratio of 21.5dB. The PZT-actuated nitride coil phase modulator operates at 493 nm with a $V\pi$ of 2.8V and low residual amplitude modulation (RAM) of -34 dB at a 1kHz offset. The bus-coupled ring resonator modulator operates at 493 nm and the add-drop ring resonator modulator operates at 780 nm. The ring-based modulators have an intrinsic quality factor ($Q_i$) of 3.4 million and 1.9 million, a linear tuning strength of 0.9 GHz/V and 1 GHz/V, and a 3-dB bandwidth of DC to 2.6 MHz and DC to 10 MHz, respectively. All four modulator designs maintain the native low optical waveguide loss of SiN, are DC coupled with broadband frequency response, operate independent of wavelength, and consume only tens of nW per actuator. Such solutions unlock the potential for further integration with other precision silicon nitride components to realize chip-scale atomic and quantum systems.

## 1. Introduction

High-efficiency integrated optical modulators are an essential component towards realizing compact and robust quantum technologies. Specifically, the on-chip miniaturization of modulators responsible for laser stabilization, disciplining to atomic transitions, and quantum functions such as gating and qubit control will pave the way for portable, robust trapped-ion [1,2] and neutral atom quantum computers [3], atomic clocks [4,5], and quantum sensors [6,7]. Given the diversity of atomic species and a wide range of architectures and protocols, it is important that these modulators can operate across visible to NIR wavelengths, that they provide a choice of different modulation formats, maintain characteristics of high-performance integrated waveguides and components, and operate with low power dissipation and ease of interface to control electronics. For example, trapped $Ba^+$ quantum computing architectures can utilize integrated phase and amplitude modulators operating at the 493 nm $6S_{1/2} \rightarrow 6P_{1/2}$ quadrupole transition for cooling and state detection [8,9]. 532 nm modulators can be used for off-resonant atom control in a variety of systems including driving Raman transitions in $Ba^+$ [10], optical dipole traps [11], and optical lattices [12]. Future space-based cold rubidium atomic clocks [13] and atom interferometers [14] will benefit from modulation at the 780 nm rubidium $D_2$ line. At the same time, such modulator solutions should be compatible with a high-performance CMOS foundry compatible photonic integration platform. Such solutions will pave the way for integration with other precision components to realize lasers, photonics, and systems-on-chip solutions for atomic and quantum applications.

The CMOS-compatible silicon nitride integration platform [15] is an important solution due to the ultra-low optical loss across the blue to NIR wavelengths [16,17] and ability to integrate precision lasers and other photonic components that operate across this range [8,18–25]. To date, stress optic ring resonator amplitude modulation based on PZT actuation of $Si_3N_4$ waveguides has been demonstrated at communications wavelengths [26]. This class of control modulator provides strong DC coupling and broadband modulation out to 10s of MHz, wavelength independent operation, and does not affect the nitride waveguide loss. These characteristics are necessary for precise atom system control functions such as laser locking [27], power routing/switching [1,28], sideband modulation, feedforward phase noise reduction [29,30], and agile phase [31] and frequency control [32]. Modulation at visible to NIR frequencies has been demonstrated using the stress-optic effect in released aluminum nitride (AlN) structures [18,33,34] and the electro-optic effect in thin film lithium niobate [35]. Progress on visible light to NIR modulators that can achieve DC-coupled broadband modulation, low residual amplitude modulation (RAM), low power consumption, and wavelength independence in a CMOS-compatible integration platform is the next critical step for integration of atomic and quantum experiments.

In this work, we report on visible to NIR high modulation efficiency photonic integrated PZT-actuated $Si_3N_4$ stress-optic amplitude and phase modulators. We demonstrate four modulator types operating at 493 nm, 532 nm, and 780 nm, as illustrated in Fig. 1, for applications to quantum computing, atomic clocks, and quantum sensing. The coil Mach-Zehnder Modulator (coil MZM) employs a 5 cm long PZT actuator in one arm, the phase modulator is a straight waveguide modulator with a 5 cm long PZT actuator, and the bus-coupled and add-drop ring resonators employ PZT actuation over the ring portion. The coil MZM operates at 532 nm with a V$\pi$ of 2.8 V, an extinction ratio (ER) of 21.5 dB, and DC-coupled 3-dB bandwidth of 0.4 MHz. The coil phase modulator operates at 493 nm with a V$\pi$ of 2.8 V, 180° phase lag bandwidth of 1 MHz, and a low residual amplitude modulation (RAM) of -34 dB at a 1kHz offset. The critically coupled bus-coupled ring resonator operates at 493 nm, has an intrinsic quality factor ($Q_i$) of 3.4M, linear tuning strength of 0.9 GHz/V, ER of 18.7 dB, and DC-coupled 3-dB bandwidth of 2.6 MHz. Finally, the add-drop ring resonator operates at 780 nm, has an intrinsic quality factor ($Q_i$) of 1.9M, linear tuning strength of 1 GHz/V, and DC-coupled 3-dB bandwidth of 10 MHz. This progress in visible light to NIR integrated amplitude and phase modulation establishes a clear path towards fully integrated quantum technologies in the silicon nitride integration platform.

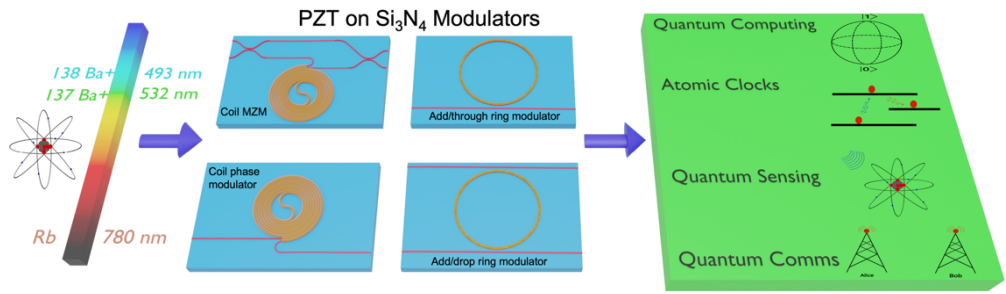

Fig. 1 Illustration of the four PZT silicon nitride integrated visible to NIR modulators (blue background chips in center) with applications to various quantum technologies (green background on right). Examples of the optical wavelength and atomic species experiments are illustrated (left) with 493 nm modulation used for the $Ba^+$ ion cooling transition and the 532 nm modulation for $Ba^+$ ion off-resonant control e.g., driving Raman transitions in $^{137}Ba^+$. The 780 nm modulation is applicable to the Rb $D_2$ cooling transition.

## 2. Results

The modulator layouts for the coil MZM, coil phase modulator, and ring resonator modulators are shown in Fig. 2. All designs utilize integrated $Si_3N_4$ waveguides with monolithically integrated PZT actuators [26,36]. The $Si_3N_4$ waveguides are 20 nm thick and 2 µm wide for the 493 nm and 532 nm devices and 120 nm thick and 900 nm wide for the 780 nm devices and are fabricated using the ultra-low loss CMOS-compatible process [15]. The integrated PZT actuator and platinum electrode fabrication [37–40] is performed after the upper cladding oxide deposition. The PZT and electrodes are offset from the waveguide by 0-5 µm, depending on the design, and are designed to achieve a high lateral strain across the nitride core for strong optical modulation while minimizing the overlap with the optical mode [26]. This back-end process avoids undercut or "pull back" structures commonly required in aluminum nitride (AlN) stress-optic modulators [34]. The PZT can be driven with up to 20 V, and all these devices register a leakage current of less than 1 nA, corresponding to a negligible power consumption of ~20nW.

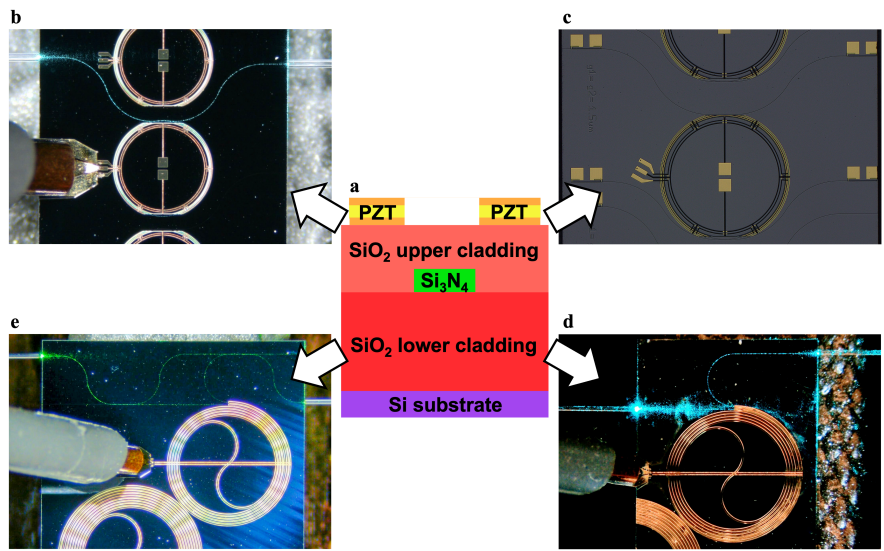

Fig. 2 **a** Cross section of the silicon nitride waveguide and back-end deposited PZT actuator. **b** Single bus add-through ring modulator under test at 493 nm featuring GSG probe to deliver

electrical signal to the PZT actuator. **c** Coil MZM under test at 532 nm. **d** Coil phase
modulator under test at 493 nm.

The coil MZM consists of a Mach-Zehnder design with a 5 cm coil PZT actuator on one of the arms as shown in Fig. 2e. By applying a voltage to the PZT, a lateral strain is induced in the waveguide, corresponding to a change in waveguide refractive index from the stress-optic effect. As a result, the phase of the light linearly shifts with a measured tuning coefficient of 716 MHz/V (0.68 pm/V), corresponding to a $V\pi$ of 2.8 V, $V\pi L$ of 14.0 V·cm, and $V\pi L\alpha$ of 6.71 V·dB. The directional coupler is designed to be an equal 50/50 split, achieving an ER of 21.5 dB at 532 nm operation. The advantages of the coil MZM are low optical losses ($\alpha = 0.24$ dB/cm), high modulation efficiency (low $V\pi$), and PZT power consumption of only 5 nW. To generate the 532 nm laser input for the coil MZM, a 1064 nm distributed Bragg reflector laser is amplified by a ytterbium doped fiber amplifier and frequency doubled down to 532 nm with periodically poled lithium niobate.

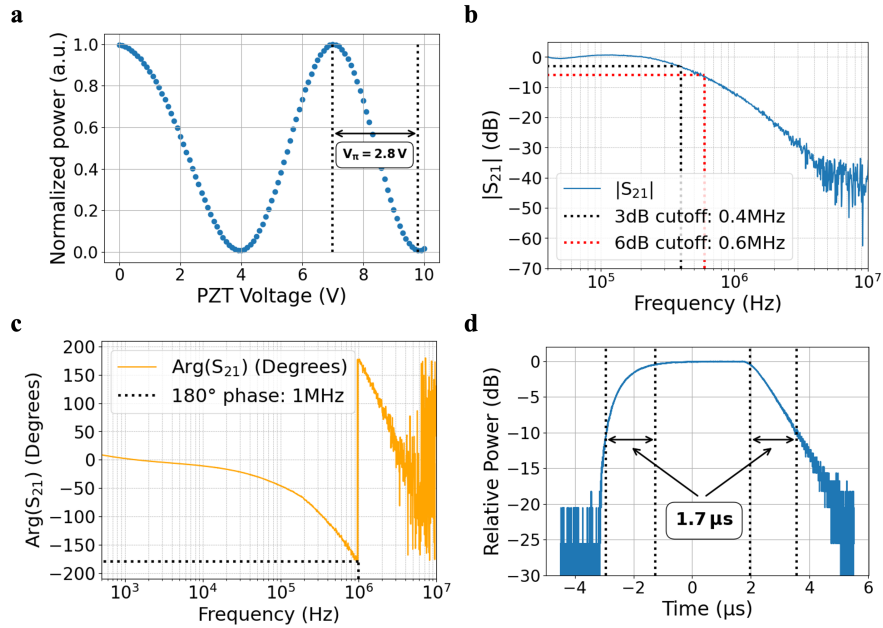

Fig. 3 Measurements of the coil MZM. **a** Static tuning of the PZT actuator. The coil MZM operates with a $V\pi$ of 2.8V and ER of 21.5 dB. **b** Amplitude response of the coil MZM showing a 3-dB bandwidth of 0.4 MHz and a 6-dB bandwidth of 0.6 MHz. **c** Phase response of the coil MZM showing a 180° phase lag point of 1MHz. **d** Optical response of the MZM to a voltage step function, showing a 90/10 rise time of 1.7 μs.

The coil MZM's modulation bandwidth is limited to ~1 MHz because the large area of the PZT actuator leads to a relatively high capacitance on the PZT of 19nF to realize a 5 cm long coil upon which the PZT actuator is deposited. Whereas ring resonator modulator bandwidths are often limited by the photon cavity lifetime, the modulation bandwidth of the coil MZM is limited only by the capacitance of the actuator, so there is a design trade-off between bandwidth and tuning efficiency. The waveguide was designed for TE0 single mode operation at 532 nm. The coil MZM exhibits a 90/10 rise time of 1.7 μs when driven with a voltage step function. The coil MZM is operational as an amplitude modulator with high extinction ratio (ER) only at wavelengths near 532 nm since that is the wavelength for which the splitting ratio of the directional coupler is closest to 50/50; however, simple waveguide design changes could produce similar modulators at any wavelength in the $Si_3N_4$ transparency window.

The coil phase modulator is a SiN waveguide with a 5 cm long actuator, realized by dicing off the input splitter of the coil MZM and using only the PZT-actuated arm (Fig. 2d). To measure the frequency response of the device, it was placed on one arm of a balanced fiber MZI to turn phase fluctuations into amplitude fluctuations and modulated at quadrature. The output of the fiber MZI was sent to a photodetector and the $S_{21}$ shown in Fig. 4a and 4b was measured using a vector network analyzer (VNA) (Agilent N5230-90017). In PDH locking schemes and other applications like optical gyroscopes [41], low-RAM modulation is desired [42]. The RAM is measured by detecting the output of the phase and viewing on a VNA. The RAM is defined as the relative level of this $|S_{21}|$ signal from the $|S_{21}|$ measured in Fig. 4a, attenuating as necessary so the carrier and RAM have the same optical power. The RAM is limited by the noise floor of the VNA and photodetector relative to the available optical power of the carrier signal. The RAM may be lower than reported here, since the design does not inherently generate any unwanted amplitude modulation as with ring resonator based phase modulators. Etalon effects and associated RAM were mitigated using index matching gel at the fiber-chip interface.

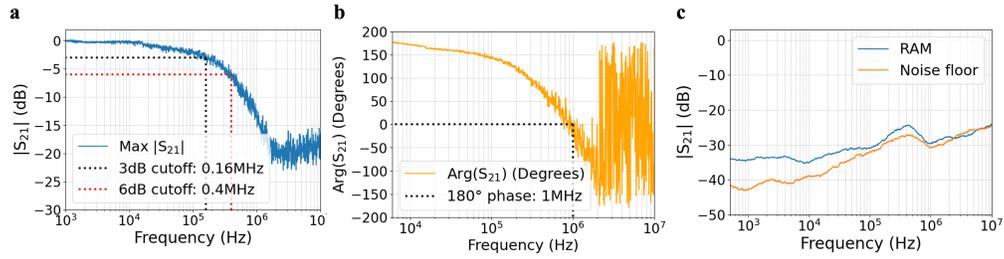

Fig. 4 Measurements of the coil phase modulator. **a** Frequency response of the coil phase modulator with 493 nm laser input. The phase modulator was placed on one arm of an off chip balanced fiber MZI to turn the phase response into an amplitude response for measurement. The maximum of 20 measurements at each frequency offset is plotted. **b** Phase of $S_{21}$ of one of the 20 measurements shown in **a**. **c** RAM measurement of the coil phase modulator with 493 nm laser input showing a RAM of -34dB at 10kHz frequency offset.

Two PZT-actuated ring resonator modulators are demonstrated. The first is a 493 nm critically coupled bus-coupled ring with 750 μm radius and a 20 nm nitride core. The device was tested using a DRS Daylight Stretto 493 nm laser. We measure an intrinsic quality factor ($Q_i$) of 3.4M, loaded quality factor ($Q_L$) of 1.9M, full width half maximum (FWHM) linewidth of 324 MHz, propagation loss of 0.24 dB/cm, and ER of 18.7 dB. The high ER of the resonator has the potential for quantum experiment gating functions. The ferroelectric nature of PZT leads to nonlinear tuning when the drive voltage exceeds roughly ±1.5V. Outside of the hysteresis regime, linear tuning is observed at 0.92 GHz/V (0.75 pm/V), corresponding to $V\pi$ of 20.7 V, $V\pi L$ of 9.75 V·cm, and $V\pi L\alpha$ of 2.34 V·dB.

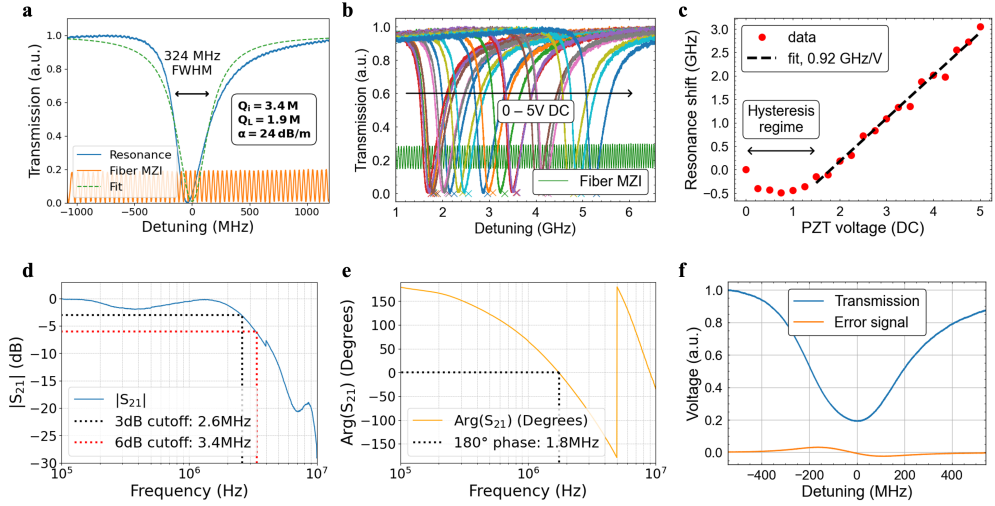

Fig. 5 Measurements of the single bus add-through 493 nm ring modulator. **a** Resonator quality factor measurement. The orange trace is a 40.2 MHz FSR fiber MZI used as a frequency ruler to measure and fit to the resonance. **b** Transmission spectra of the resonator for applied DC voltages to the PZT electrode every 0.25 V from 0 V to 5 V. **c** Linearity of static PZT actuator tuning. Excluding the 0 – 1.5 V region of hysteresis, a linear tuning of 0.92 GHz/V is observed. **d** Amplitude of $S_{21}$ frequency response as measured by a VNA. **e** Phase response of the $S_{21}$ measurement. **f** By applying a 1 MHz small signal modulation to the PZT actuator, an error signal that can be used for PDH locking is observed, and the resonance broadens due to the sidebands.

Tuning the laser to the quadrature point of the resonator and biasing the PZT electrode above its region of hysteresis, small-signal modulation is applied to the device and the response measured using a VNA. A 3-dB modulation bandwidth is measured to be DC to 2.6 MHz and 6-dB modulation bandwidth to be DC to 3.4 MHz with a $180°$ phase lag point of 1.8 MHz. The bandwidth of the modulator is limited by the capacitance of the PZT actuator and the photon lifetime in the ring resonator, which motivates the use of a ring modulator with only moderate Q [26]. These ring modulators can also be used as a phase modulator by tuning the laser to the resonance minimum and actuating with small signal modulation. This is demonstrated in Fig. 5f. to generate an error signal that can be used for PDH locking.

The second type of ring modulator is a 750 μm radius add-drop resonator modulator. The resonator is slightly under-coupled at 780 nm, achieving an ER of 12.1 dB, and achieves a $Q_i$ of 1.9M, $Q_L$ of 0.7M, FWHM linewidth of 539 MHz, and propagation loss of 0.27 dB/cm (Fig. 6). Outside of the hysteresis regime, linear tuning is observed at 1.01 GHz/V (2.1 pm/V), corresponding to $V\pi$ of $18.8$ V, $V\pi L$ of $8.86$ V·cm, and $V\pi L\alpha$ of $2.39$ V·dB. Both resonators also have thermal tuners if additional slow tuning is required for more advanced locking and modulation schemes.

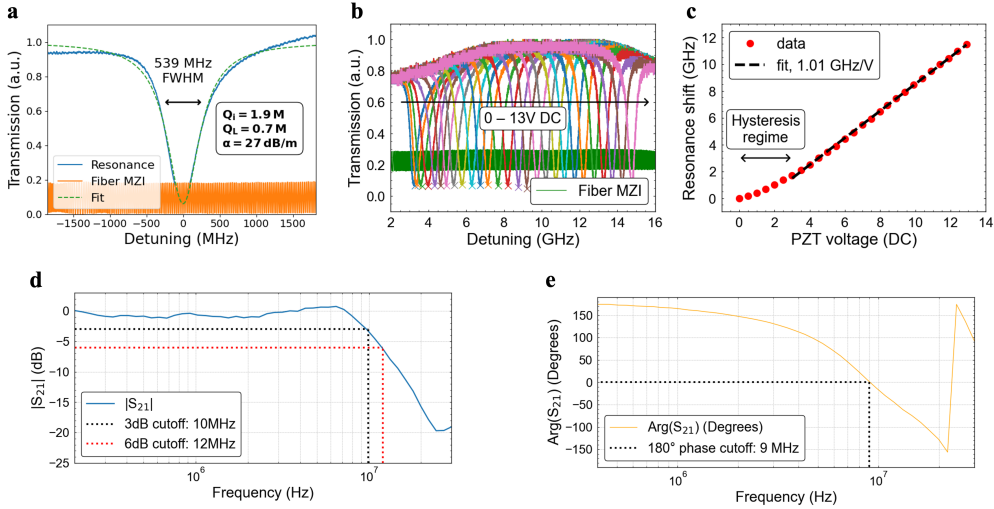

Fig. 6 Measurements of two-bus add-drop 780 nm ring modulator. **a** Resonator Q measurement. The orange trace is a 24.8 MHz FSR fiber MZI used as a frequency ruler to measure and fit to the resonance. **b** Transmission spectra of the resonator for applied DC voltages to the PZT electrode every 0.5 V from 0 V to 13 V. **c** Linearity of static PZT actuator tuning. Excluding the 0 – 3 V region of hysteresis, a linear tuning of 1.01 GHz/V is observed. **d** Amplitude of $S_{21}$ frequency response measured by a Red Pitaya. **e** Phase response of the $S_{21}$ measurement.

## 3. Discussion

We report a class of versatile photonic integrated PZT-actuated stress-optic modulators capable of operation over the 493 to 780 nm VIS to NIR range. We demonstrate Mach-Zehnder interferometer, pure phase, and bus-ring and add-drop modulator designs with V$\pi$ as low as 2.8 V, DC-coupled 3-dB bandwidth up to 10 MHz, on-off ER as high as 21.5 dB, and only 10s of nWs of electrical power consumption. The PZT actuator maintains the ultra-low silicon nitride waveguide loss and ultra-high Q, and provides a path to integration with other silicon nitride photonic integrated components including lasers for external amplitude or phase modulation and direct agile frequency tuning [19]. The bandwidth of PZT silicon nitride actuators has been demonstrated up to 70 MHz [36] and are well suited to atomic quantum control functions and protocols such as polarization gradient cooling frequency control sequences [32], Zeeman qubits [43], and PDH locking integrated lasers to integrated coil references [44]. Additionally, the strong DC stress-optic coefficient of PZT provides stronger quiescent set points and tuning that is orders of magnitude stronger than other stress-optic actuators [45]. The measured V$\pi$L$\alpha$ values of 6.71 V·dB for the coil MZM and 2.34 V·dB for the 493 nm ring modulator have the potential in the future for sub-volt levels with a further reduction in waveguide losses [16,46]. These modulators can find applications in scaling quantum technologies such as optical tweezers [47,48], trapped-ions [49], and non-Abelian anyons [50]. These results provide a clear path to integration of modulators with other visible $Si_3N_4$ components for fully integrated cold atom and trapped-ion systems on-chip.

**Funding.** This work is funded in part by grants from DEVCOM Army Research Laboratory award number W911NF-22-2-0056), U.S. Army Research Office award number W911NF2310179), DARPA GRYPHON award number HR0011-22-2-0008, and a Quantinuum gift.

**Acknowledgment.** We thank DRS Daylight Solutions for their loan of the 493 nm Stretto laser and Chuan Zhong for dicing the chips. Jiawei Wang and Nitesh Chauhan's work was performed while they were at UCSB and affiliations 2-4 are their current affiliations.

**Disclosures.** D.J.B.'s work was funded by ColdQuanta d.b.a. Infleqtion. D.J.B. has consulted for Infleqtion, received compensation, and owns stock.

**Data availability.** Data underlying the results presented in this paper are not publicly available at this time but may be obtained from the authors upon reasonable request.

## References

1. C. W. Hogle, D. Dominguez, M. Dong, A. Leenheer, H. J. McGuinness, B. P. Ruzic, M. Eichenfield, and D. Stick, "High-fidelity trapped-ion qubit operations with scalable photonic modulators," npj Quantum Inf **9**, 1–6 (2023).
2. H. Häffner, C. F. Roos, and R. Blatt, "Quantum computing with trapped ions," Physics Reports **469**, 155–203 (2008).
3. T. M. Graham, Y. Song, J. Scott, C. Poole, L. Phuttitarn, K. Jooya, P. Eichler, X. Jiang, A. Marra, B. Grinkemeyer, M. Kwon, M. Ebert, J. Cherek, M. T. Lichtman, M. Gillette, J. Gilbert, D. Bowman, T. Ballance, C. Campbell, E. D. Dahl, O. Crawford, N. S. Blunt, B. Rogers, T. Noel, and M. Saffman, "Multi-qubit entanglement and algorithms on a neutral-atom quantum computer," Nature **604**, 457–462 (2022).
4. A. D. Ludlow, M. M. Boyd, J. Ye, E. Peik, and P. O. Schmidt, "Optical atomic clocks," Rev. Mod. Phys. **87**, 637–701 (2015).
5. Z. L. Newman, V. Maurice, T. Drake, J. R. Stone, T. C. Briles, D. T. Spencer, C. Fredrick, Q. Li, D. Westly, B. R. Ilic, B. Shen, M.-G. Suh, K. Y. Yang, C. Johnson, D. M. S. Johnson, L. Hollberg, K. J. Vahala, K. Srinivasan, S. A. Diddams, J. Kitching, S. B. Papp, and M. T. Hummon, "Architecture for the photonic integration of an optical atomic clock," Optica, OPTICA **6**, 680–685 (2019).
6. A. Kodigala, M. Gehl, G. W. Hoth, J. Lee, C. T. DeRose, A. Pomerene, C. Dallo, D. Trotter, A. L. Starbuck, G. Biedermann, P. D. D. Schwindt, and A. L. Lentine, "High-performance silicon photonic single-sideband modulators for cold-atom interferometry," Science Advances **10**, eade4454 (2024).
7. H. S. Stokowski, T. P. McKenna, T. Park, A. Y. Hwang, D. J. Dean, O. T. Celik, V. Ansari, M. M. Fejer, and A. H. Safavi-Naeini, "Integrated quantum optical phase sensor in thin film lithium niobate," Nat Commun **14**, 3355 (2023).
8. A. A. Savchenkov, J. E. Christensen, D. Hucul, W. C. Campbell, E. R. Hudson, S. Williams, and A. B. Matsko, "Application of a self-injection locked cyan laser for Barium ion cooling and spectroscopy," Sci Rep **10**, 16494 (2020).
9. A. Binai-Motlagh, M. Day, N. Videnov, N. Greenberg, C. Senko, and R. Islam, "A guided light system for agile individual addressing of Ba$^+$ qubits with $10^{-4}$ level intensity crosstalk," (2023).
10. A. S. Sotirova, B. Sun, J. D. Leppard, A. Wang, M. Wang, A. Vazquez-Brennan, D. P. Nadlinger, S. Moser, A. Jesacher, C. He, F. Pokorny, M. J. Booth, and C. J. Ballance, "Low Cross-Talk Optical Addressing of Trapped-Ion Qubits Using a Novel Integrated Photonic Chip," (2023).
11. A. Jenkins, J. W. Lis, A. Senoo, W. F. McGrew, and A. M. Kaufman, "Ytterbium Nuclear-Spin Qubits in an Optical Tweezer Array," Phys. Rev. X **12**, 021027 (2022).
12. G.-B. Jo, J. Guzman, C. K. Thomas, P. Hosur, A. Vishwanath, and D. M. Stamper-Kurn, "Ultracold Atoms in a Tunable Optical Kagome Lattice," Phys. Rev. Lett. **108**, 045305 (2012).
13. L. Liu, D.-S. Lü, W.-B. Chen, T. Li, Q.-Z. Qu, B. Wang, L. Li, W. Ren, Z.-R. Dong, J.-B. Zhao, W.-B. Xia, X. Zhao, J.-W. Ji, M.-F. Ye, Y.-G. Sun, Y.-Y. Yao, D. Song, Z.-G. Liang, S.-J. Hu, D.-H. Yu, X. Hou, W. Shi, H.-G. Zang, J.-F. Xiang, X.-K. Peng, and Y.-Z. Wang, "In-orbit operation of an atomic clock based on laser-cooled 87Rb atoms," Nat Commun **9**, 2760 (2018).
14. K. Bongs, M. Holynski, J. Vovrosh, P. Bouyer, G. Condon, E. Rasel, C. Schubert, W. P. Schleich, and A. Roura, "Taking atom interferometric quantum sensors from the laboratory to real-world applications," Nat Rev Phys **1**, 731–739 (2019).
15. J. F. Bauters, M. J. R. Heck, D. John, D. Dai, M.-C. Tien, J. S. Barton, A. Leinse, R. G. Heideman, D. J. Blumenthal, and J. E. Bowers, "Ultra-low-loss high-aspect-ratio $Si_3N_4$ waveguides," Opt. Express, OE **19**, 3163–3174 (2011).
16. N. Chauhan, J. Wang, D. Bose, K. Liu, R. L. Compton, C. Fertig, C. W. Hoyt, and D. J. Blumenthal, "Ultra-low loss visible light waveguides for integrated atomic, molecular, and quantum photonics," Opt. Express, OE **30**, 6960–6969 (2022).
17. K. Liu, N. Jin, H. Cheng, N. Chauhan, M. W. Puckett, K. D. Nelson, R. O. Behunin, P. T. Rakich, and D. J. Blumenthal, "Ultralow 0.034 dB/m loss wafer-scale integrated photonics realizing 720 million Q and 380 μW threshold Brillouin lasing," Opt. Lett., OL **47**, 1855–1858 (2022).
18. A. Siddharth, A. B. Gardner, X. Ji, S. U. Hulyal, M. S. Reichler, A. Attanasio, J. Riemensberger, S. A. Bhave, N. Volet, S. Bianconi, and T. J. Kippenberg, "Narrow-linewidth, piezoelectrically tunable photonic integrated blue laser," (2025).
19. A. Isichenko, N. Chauhan, K. Liu, M. W. Harrington, and D. J. Blumenthal, "Chip-Scale, Sub-Hz Fundamental Sub-kHz Integral Linewidth 780 nm Laser through Self-Injection-Locking a Fabry-Pérot laser to an Ultra-High Q Integrated Resonator," Sci Rep **14**, 27015 (2024).

20. H. Nejadriahi, E. Kittlaus, D. Bose, N. Chauhan, J. Wang, M. Fradet, M. Bagheri, A. Isichenko, D. Heim, S. Forouhar, and D. J. Blumenthal, "Sub-100 Hz intrinsic linewidth 852 nm silicon nitride external cavity laser," Opt. Lett., OL **49**, 7254–7257 (2024).
21. N. Chauhan, A. Isichenko, K. Liu, J. Wang, Q. Zhao, R. O. Behunin, P. T. Rakich, A. M. Jayich, C. Fertig, C. W. Hoyt, and D. J. Blumenthal, "Visible light photonic integrated Brillouin laser," Nat Commun **12**, 4685 (2021).
22. N. Chauhan, C. Caron, J. Wang, A. Isichenko, N. Helaly, K. Liu, R. J. Niffenegger, and D. J. Blumenthal, "Trapped ion qubit and clock operations with a visible wavelength photonic coil resonator stabilized integrated Brillouin laser," (2024).
23. M. W. Puckett, K. Liu, N. Chauhan, Q. Zhao, N. Jin, H. Cheng, J. Wu, R. O. Behunin, P. T. Rakich, K. D. Nelson, and D. J. Blumenthal, "422 Million intrinsic quality factor planar integrated all-waveguide resonator with sub-MHz linewidth," Nat Commun **12**, 934 (2021).
24. A. Isichenko, N. Chauhan, D. Bose, J. Wang, P. D. Kunz, and D. J. Blumenthal, "Photonic integrated beam delivery for a rubidium 3D magneto-optical trap," Nature communications **14**, 3080 (2023).
25. J. K. S. Poon, A. Govdeli, A. Sharma, X. Mu, F.-D. Chen, T. Xue, and T. Liu, "Silicon photonics for the visible and near-infrared spectrum," Adv. Opt. Photon., AOP **16**, 1–59 (2024).
26. J. Wang, K. Liu, M. W. Harrington, R. Q. Rudy, and D. J. Blumenthal, "Silicon nitride stress-optic microresonator modulator for optical control applications," Opt. Express, OE **30**, 31816–31827 (2022).
27. J. I. Thorpe, K. Numata, and J. Livas, "Laser frequency stabilization and control through offset sideband locking to optical cavities," Opt. Express, OE **16**, 15980–15990 (2008).
28. A. J. Menssen, A. Hermans, I. Christen, T. Propson, C. Li, A. J. Leenheer, M. Zimmermann, M. Dong, H. Larocque, H. Raniwala, G. Gilbert, M. Eichenfield, and D. R. Englund, "Scalable photonic integrated circuits for high-fidelity light control," Optica, OPTICA **10**, 1366–1372 (2023).
29. T. Denecker, Y. T. Chew, O. Guillemant, G. Watanabe, T. Tomita, K. Ohmori, and S. De Léséleuc, "Measurement and feedforward correction of the fast phase noise of lasers," Phys. Rev. A **111**, 042614 (2025).
30. L. Yan, "High-Power Clock Laser Spectrally Tailored for High-Fidelity Quantum State Engineering," Phys. Rev. X **15**, (2025).
31. C. LeDesma, K. Mehling, J. D. Wilson, M. Nicotra, and M. Holland, "Universal gate set for optical lattice based atom interferometry," Phys. Rev. Research **7**, 013246 (2025).
32. A. Isichenko, S. Carpenter, N. Montifiore, J. Wang, M. Dangi, N. Chauhan, P. Mukherjee, X. Yang, N. Indukuri, M. W. Harrington, C. Zhong, I. M. Kierzewski, R. Q. Rudy, J. T. Choy, and D. J. Blumenthal, "Sub-Doppler rubidium atom cooling using a programmable agile integrated PZT-on-SiN resonator," (2026).
33. M. Dong, D. Heim, A. Witte, G. Clark, A. J. Leenheer, D. Dominguez, M. Zimmermann, Y. H. Wen, G. Gilbert, D. Englund, and M. Eichenfield, "Piezo-optomechanical cantilever modulators for VLSI visible photonics," APL Photonics **7**, 051304 (2022).
34. P. R. Stanfield, A. J. Leenheer, C. P. Michael, R. Sims, and M. Eichenfield, "CMOS-compatible, piezo-optomechanically tunable photonics for visible wavelengths and cryogenic temperatures," Opt. Express, OE **27**, 28588–28605 (2019).
35. N. Achuthan, X. Li, L. Magalhães, X. Zhu, H. Warner, X. Zuo, S. W. Ding, K. Powell, D. Renaud, D. Assumpcao, M. Pushkarsky, A. Shams-Ansari, and M. Lončar, "Broadband electro-optic modulation at blue wavelengths in thin-film lithium niobate," in *Photonics for Quantum 2025* (SPIE, 2025), Vol. 13563, p. 1356302.
36. M. W. Harrington, S. M. Zhu, R. Chawlani, K. Liu, S. Sun, R. Liu, X. Yi, and D. J. Blumenthal, "Integrated Architecture for the Automated Generation and Coil Stabilization of a PZT-Enabled Microcomb," (2025).
37. J. S. Pulskamp, S. S. Bedair, R. G. Polcawich, C. D. Meyer, I. Kierzewski, and B. Maack, "High-Q and Capacitance Ratio Multilayer Metal-on-Piezoelectric RF MEMS Varactors," IEEE Electron Device Letters **35**, 871–873 (2014).
38. S. S. Bedair, J. S. Pulskamp, C. D. Meyer, M. Mirabelli, R. G. Polcawich, and B. Morgan, "High-Performance Micromachined Inductors Tunable by Lead Zirconate Titanate Actuators," IEEE Electron Device Letters **33**, 1483–1485 (2012).
39. G. L. Smith, J. S. Pulskamp, L. M. Sanchez, D. M. Potrepka, R. M. Proie, T. G. Ivanov, R. Q. Rudy, W. D. Nothwang, S. S. Bedair, C. D. Meyer, and R. G. Polcawich, "PZT-Based Piezoelectric MEMS Technology," Journal of the American Ceramic Society **95**, 1777–1792 (2012).
40. J. S. Pulskamp, R. G. Polcawich, R. Q. Rudy, S. S. Bedair, R. M. Proie, T. Ivanov, and G. L. Smith, "Piezoelectric PZT MEMS technologies for small-scale robotics and RF applications," MRS Bulletin **37**, 1062–1070 (2012).
41. D. A. Pogorelaya, M. A. Smolovik, V. E. Strigalev, A. S. Aleynik, and I. G. Deyneka, "An investigation of the influence of residual amplitude modulation in phase electro-optic modulator on the signal of fiber-optic gyroscope," J. Phys.: Conf. Ser. **735**, 012040 (2016).
42. W. Zhang, M. J. Martin, C. Benko, J. L. Hall, J. Ye, C. Hagemann, T. Legero, U. Sterr, F. Riehle, G. D. Cole, and M. Aspelmeyer, "Reduction of residual amplitude modulation to $1 \times 10^{-6}$ for frequency modulation and laser stabilization," Opt. Lett. **39**, 1980 (2014).
43. T. Ruster, C. T. Schmiegelow, H. Kaufmann, C. Warschburger, F. Schmidt-Kaler, and U. G. Poschinger, "A long-lived Zeeman trapped-ion qubit," Appl. Phys. B **122**, 254 (2016).

44. M. Song, N. Chauhan, M. W. Harrington, N. Montifiore, K. Liu, A. S. Hunter, C. Caron, A. Isichenko, R. J. Niffenegger, and D. J. Blumenthal, "Octave Spanning Visible to SWIR Integrated Coil-Stabilized Brillouin Lasers," (2025).
45. A. Mazzalai, D. Balma, N. Chidambaram, R. Matloub, and P. Muralt, "Characterization and Fatigue of the Converse Piezoelectric Effect in PZT Films for MEMS Applications," Journal of Microelectromechanical Systems **24**, 831–838 (2015).
46. N. Montifiore, A. Isichenko, J. Wang, N. Chauhan, M. W. Harrington, M. Pushkarsky, and D. J. Blumenthal, "Integrated Low-Power Blue Light PZT Silicon Nitride Ring Modulator for Atomic and Quantum Applications," in *CLEO 2025 (2025), Paper SS160_5* (Optica Publishing Group, 2025), p. SS160_5.
47. N.-C. Chiu, E. C. Trapp, J. Guo, M. H. Abobeih, L. M. Stewart, S. Hollerith, P. L. Stroganov, M. Kalinowski, A. A. Geim, S. J. Evered, S. H. Li, X. Lyu, L. M. Peters, D. Bluvstein, T. T. Wang, M. Greiner, V. Vuletić, and M. D. Lukin, "Continuous operation of a coherent 3,000-qubit system," Nature 1–3 (2025).
48. H. J. Manetsch, G. Nomura, E. Bataille, X. Lv, K. H. Leung, and M. Endres, "A tweezer array with 6100 highly coherent atomic qubits," Nature 1–3 (2025).
49. S.-A. Guo, Y.-K. Wu, J. Ye, L. Zhang, W.-Q. Lian, R. Yao, Y. Wang, R.-Y. Yan, Y.-J. Yi, Y.-L. Xu, B.-W. Li, Y.-H. Hou, Y.-Z. Xu, W.-X. Guo, C. Zhang, B.-X. Qi, Z.-C. Zhou, L. He, and L.-M. Duan, "A site-resolved two-dimensional quantum simulator with hundreds of trapped ions," Nature **630**, 613–618 (2024).
50. S. Sun, X. Wang, S. Li, D. Zhang, Q.-D. Chen, and X.-L. Zhang, "Reconfigurable non-Abelian integrated photonics," Nat Commun **16**, 7089 (2025).